\documentclass{article}

\usepackage{PRIMEarxiv}
\usepackage{amsmath}
\usepackage{multirow}

\usepackage[utf8]{inputenc} 
\usepackage[T1]{fontenc}    
\usepackage{hyperref}       
\usepackage{url}            
\usepackage{booktabs}       
\usepackage{amsfonts}       
\usepackage{nicefrac}       
\usepackage{microtype}      
\usepackage{lipsum}
\usepackage{fancyhdr}       
\usepackage{graphicx}       
\graphicspath{{media/}}     

\title{STUDENT PAPER: Preliminary Results of Applying Modified MSA Algorithm on Quantum Annealers (MAQ)
\thanks{\textit{\underline{Citation}}: 
\textbf{Lee, M., STUDENT PAPER: Preliminary Results of Applying Modified MSA Algorithm on Quantum Annealers (MAQ),
https://doi.org/10.22369/issn.2153-4136/14/1/5}} 
}

\author{
  Melody Lee \\
   \\
  The North Carolina School of Science and Mathematics \\
  Durham, NC\\
  lee23m@ncssm.edu
}

\begin{document}
\maketitle

\begin{abstract}
We propose a modified MSA algorithm on quantum annealers with applications in areas of bioinformatics and genetic sequencing. To understand the human genome, researchers compare extensive sets of these genetic sequences -- or their protein counterparts -- to identify patterns. This comparison begins with the alignment of the set of (multiple) sequences. However, this alignment problem is considered nondeterministically-polynomial time complete and, thus, current classical algorithms at best rely on brute force or heuristic methods to find solutions. Quantum annealing algorithms are able to bypass this need for sheer brute force due to their use of quantum mechanical properties. However, due to the novelty of these algorithms, many are rudimentary in nature and limited by hardware restrictions. We apply progressive alignment techniques to modify annealing algorithms, achieving a linear reduction in spin usage whilst introducing more complex heuristics to the algorithm. This opens the door for further exploration into quantum computing-based bioinformatics, potentially allowing for a deeper understanding of disease detection and monitoring.
\end{abstract}

\section{Introduction}

\subsection{Alignments in Disease Detection and Prevention}
In a single year, over 850 million years of healthy life may be lost to disease and disabilities \cite{owidburdenofdisease}. In fact, an estimated 50\% of the United States population is living with a chronic disease \cite{holman_2020}. Presently, there is an inadequate response to this healthcare crisis, as most of the attention in epidemiological research and health care has been centered around acute diseases \cite{holman_2020}. Human genomes are explicit factors in determining susceptibility to some of these diseases \cite{klebanov_2018}. Comparison of their constituent genetic sequences may one day reveal knowledge that permits for early diagnosis or monitoring of heritable diseases of at-risk individuals \cite{huang_shah_yao_2019}. Furthermore, the comparison of sequences has heavy bearing on treatment procedures as well. For instance, the study of large sets of DNA sequences can allow researchers to eventually predict patient response to chemotherapy or other treatments \cite{phillips_trosman_kelley_pletcher_douglas_weldon_2014}. This could allow for the determination of personalized treatment options for patients in order to maximize their chances of recovery, including cancer. Therefore, emphasis on the comparison of the genetics underlying major actors in these diseases -- from protein mutations to patient genomes -- is needed. 

There is a clear issue, however. The sheer length of genetic sequences is comparable to the circumference of the Earth or even the distance to the moon. Each genetic sequence can contain on the magnitude of several thousand base pairs. Analysis of large sets of these sequences consumes significant computing resources. Protein sequences are no better, with the length of the amino acid sequences on a similar order of magnitude.  In spite of the limited alphabet these sequences are composed of -- pulling from sets of a mere four base pairs or twenty amino acids -- these sequences are responsible for the behavior of countless diseases in existence, and thus researchers have sought various methods of analyzing them. 

\begin{table}[htb]
\centering
\caption{An example alignment for a set of three genetic sequences.}
\begin{tabular}{llllll}
A & T & G & - & T & T \\
A & T & - & C & T & T \\
T & T & G & C & T & -
\end{tabular}
\label{sample}
\end{table}

To compare these sequences effectively, an ideal alignment of the sequences must be found, in which gaps or shifts in the sequences are inserted to minimize the differences in each column of Table \ref{sample}. After all, it would do no good if subsequences that encode for different biological components are mistakenly compared against one another. The problem of finding the multiple sequence alignment (MSA) is an applied form of the mathematical consensus string problem \cite{sim_park_2003}. The solution seeks to find an alignment where the distances between sequences are minimized. This distance is a quantitative measurement of how well sequences are aligned, comparable to the aforementioned number of differences in each column \cite{hosangadi_2012}. For every alignment of a pair of sequences, the elements in corresponding positions are compared. The greater the discrepancies across the positions, the greater the distance between the two sequences \cite{hosangadi_2012}. This problem is analogous to finding the smallest distance between some set of locations. The given set of locations are the sequences, and their distances are the differences between each plausible alignment. The nature of this problem, therefore, centers on distance minimization, deeming it an optimization problem. While the alignment of, say, ten or twenty elements per sequence is not difficult, solving the problem for larger and larger scales can become unmanageable for the standard human mind. 

This paper functions as a simultaneous investigation into MSA algorithms and certain alterations to these algorithms that may be made. We begin by discussing existing algorithms for MSA, alongside shortcomings in the computational tools currently in use. We then transition to discussion of quantum annealing, prior to discussing our classical-inspired modifications to a MSA quantum annealing algorithm. This modified algorithm is then used to align a sample dataset and its results analyzed. 

\subsection{Existing Algorithms for MSA}
Rather than arbitrarily align these sequences, MSA algorithms systematically align sequences. While there are numerous algorithms in existence, the most common are based on the Needleman-Wunsch algorithm. This algorithm was first described in 1970 by its namesake researchers, Saul B. Needleman and Christian D. Wunsch \cite{needleman_wunsch_1970}.The initial algorithm aligns a pair of protein sequences using iterative comparison of each individual amino acid \cite{needleman_wunsch_1970}. While effective, there lies a major issue in the resource requirements for the problem. The MSA problem has non-deterministic polynomial-time hardness (NP-hardness) \cite{sim_park_2003}. As the size of the input size increases, the amount of time and memory required to find the perfect solution increases at unmanageable rates. This becomes a hindrance in effective application. Finding the ideal alignment for inputs on the same magnitude as protein or genetic sequences can take decades to process. Thus, better algorithms capable of handling larger inputs are being sought.

Over the years, researchers have developed more complex methods of computation to raise the ceiling on the size of the inputs that may be reasonably handled. Algorithms capable of running on multiple computer cores in parallel have been developed. This is analogous to having multiple people brainstorm ideas for a project, as opposed to a singular "brain" working on the task. The approach has approximately a 60\% reduction in execution time from experimental results, showing parallel processing has strong potential \cite{muhamad_ahmad_asi_murad_2018}.  

Another common -- but effective -- method aligns smaller subsets of the sequences before merging the final solution. These methods are generally categorized into two types: (1) progressive alignments and (2) iterative alignments \cite{8900740}. Progressive algorithms organize sequences based on similarity and arrange subsets of these sequences. In some cases, the sequences are arranged in a tree-like structure, such that only a few sets of parents and their children are aligned at once, reducing the load on the computer at any single point \cite{8900740}. Iterative algorithms, on the other hand, go through multiple iterations of aligning and then re-aligning sequences in overlapping subsets. Both types may use heuristics to estimate the pairwise distances between sequences prior to arrangement, allowing them to introduce reasonable steps that increase the scalability of the resultant process \cite{wang_wu_cai_2018}. 

\subsection{A Tool for Problem Solving: Quantum Computers}
While existing methods are effective effective, they are still bound by the binary nature of computing units. That is, standard classical computers have bit values restricted to either 0 and 1, or True and False, and therefore are only able to represent one state at a time \cite{steane_1997}. Quantum computers -- which make use of parallel processing and quantum mechanical properties to bypass these restrictions -- have emerged as new contenders for finding alignments \cite{zheng_qiu_gruska_2017}. 

\begin{figure}
    \centering
    \includegraphics[scale=0.20]{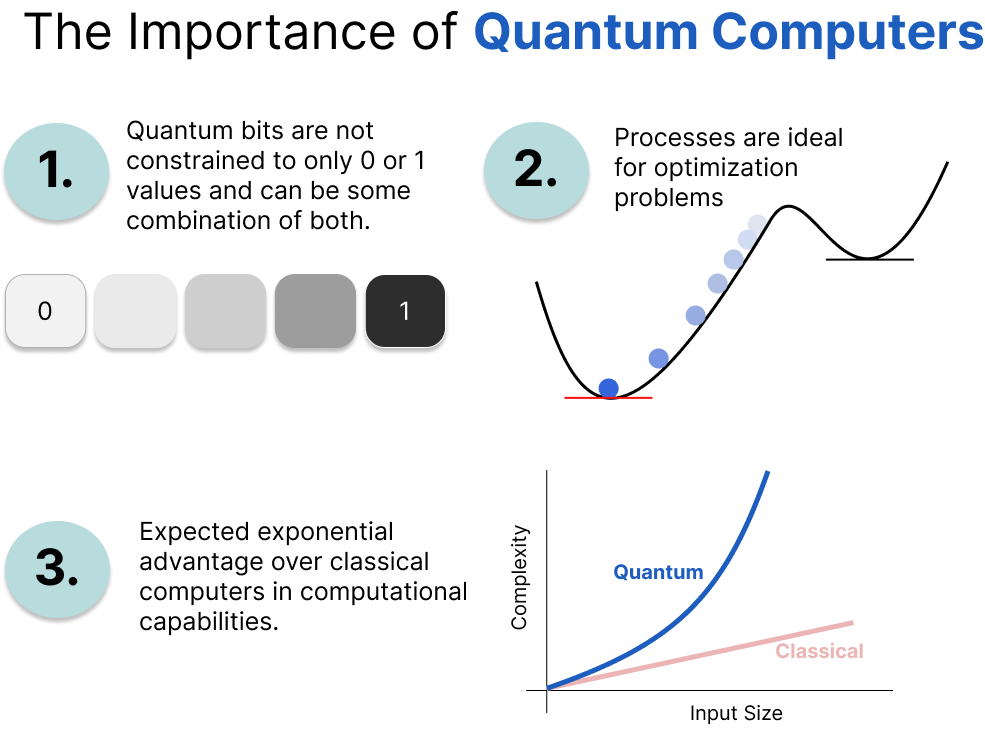}
    \caption{There exist several key characteristics of quantum computers that make them especially of interest when it comes to algorithms (created by author).}
    \label{qc_importance}
\end{figure}

While the absolute supremacy of quantum computers over their classical counterparts is yet unproven \cite{tang_2019}, they have two key properties whose partnership make computation on quantum systems especially advantageous: one, parallel processing and, two, entanglement. The parallel processing capabilities come from the ability for the quantum bits to be in a probabilistic suspension between the bit values, or in a superimposed state \cite{zheng_qiu_gruska_2017}. This phenomenon allows for an exponential number of solutions to be simultaneously represented \cite{steane_1997}. This cooperates with the second property, entanglement, to make quantum computers especially unique. The values of the quantum bits -- including those in superposition -- may be "tangled" together, such that knowledge of the value of one qubit will reveal information about other entangled qubits in the system \cite{steane_1997}. This permits for added levels of complexity \cite{steane_1997}. The combination of these quantum mechanical properties in computing makes quantum computing especially well-suited for solving NP-hard problems (Figure \ref{qc_importance}). 

For example, certain algorithms have used a combination of both classical techniques and quantum computer capabilities. Researchers have applied machine learning models to reduce the amount of memory required to store comparisons of the sequences \cite{ventura_martinez_1998}. Others have taken inspiration from the quantum mechanical properties outright in developing quantum-inspired heuristics to find alignments \cite{8900740}. 

\subsubsection{Quantum Annealing Algorithms}
\begin{figure}
    \centering
    \includegraphics[scale=0.40]{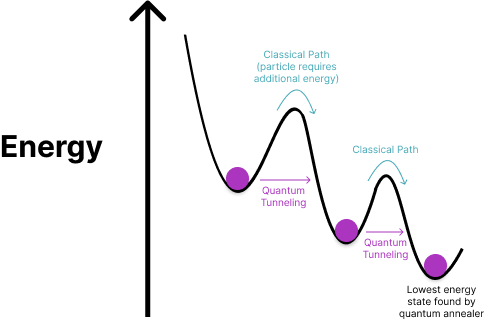}
    \caption{Quantum annealers use quantum tunneling to find the lowest energy state for the given problem space (created by author).}
    \label{annealing}
\end{figure}

Other algorithms focus on a subtype of quantum computing: quantum annealing. Quantum annealers, also known as adiabatic quantum computers, take advantage of the natural tendency for physical systems to seek out the lowest energy configurations \cite{d-wave}. A commonly used analogy to illustrate the workings of a quantum annealer involves finding the lowest point of elevation among a series of hills and valleys \cite{d-wave}. This region is analogous to the problem space defined. Classical computers find the solution to the problem by sending a singular traveler to begin at some arbitrary point in the area. This traveler finds the minimum by walking some direction determined by the classical algorithm until a local lowest point is reached. To ensure this is the absolute lowest point, the classical algorithm then proceeds to drop the traveler off again at several other locations across the area. Quantum annealers, on the other hand, bypass this repeated traversal. Rather, superposition permits for the traveler to exist simultaneously in different locations, cutting down significantly on the costliness of traversal \cite{d-wave}. To find the absolute minimum, quantum tunneling -- a phenomenon unique to particles on the quantum scale -- allows this traveler to "tunnel" directly through hills to reach the absolute minimum, rather than have to metaphorically climb all the way back up the hill (Figure \ref{annealing}). Since aligning a set of data involves minimizing the distance between each pair of sequences -- a textbook optimization problem -- the MSA problem fits neatly into the functionality of quantum annealers \cite{lindvall_2019}. 

\subsubsection{Current Shortcomings}
In spite of the potential advantage using quantum algorithms to find alignments may provide, there exist two major areas in need of immediate improvement. First, due to the relatively new nature of quantum annealers, existing algorithms tend to at best mirror rudimentary classical algorithms. That is, some algorithms mimic brute-force processes without the inclusion of more complex heuristics that aid the process, such as progressive or iterative techniques \cite{lindvall_2019}.

Secondly, modern quantum algorithms are constrained by hardware limitations \cite{fellous-asiani_chai_whitney_auffèves_ng_2021}. The reliance on the quantum properties of particles leaves the qubits susceptible to slight changes in the environment \cite{chuang_laflamme_shor_zurek_1995}. These errors result in inconsistencies between the simulated solution and experimental results returned \cite{johnstun_van_huele_2021}. Furthermore, the number of quantum nodes available for public use is restricted, largely due to the limited size of existing computers. For example, the D-Wave quantum annealer Advantage, contains just over 5000 quantum bits \cite{willsch_willsch_gonzalezcalaza_jin_deraedt_svensson_michielsen_2022} -- barely meeting current supercomputing capabilities, and there exist few available annealers larger in size. This places an upper bound on the size of the test data. Thus issues are raised. The input datasets of genetic and protein sequences are large in both size and sequence length. So, a sufficient amount of qubit spin usage in these quantum computers is needed. The development of a more efficient tool for MSA capable of bypassing the constraints of hardware limitations is needed. 

\section{Methods}
We took inspiration from classical algorithms that utilize clustering methodologies \cite{wang_wu_cai_2018}, where sequences are grouped before being progressively processed via the alignment algorithm, providing a close approximation of the solution \cite{feng_doolittle_1987}. In short, we introduced classical-inspired processing methods to the quantum annealing process. To do so, we implemented an overarching progressive alignment structure throughout the algorithm.

\begin{figure}[h]
    \includegraphics[width=\linewidth]{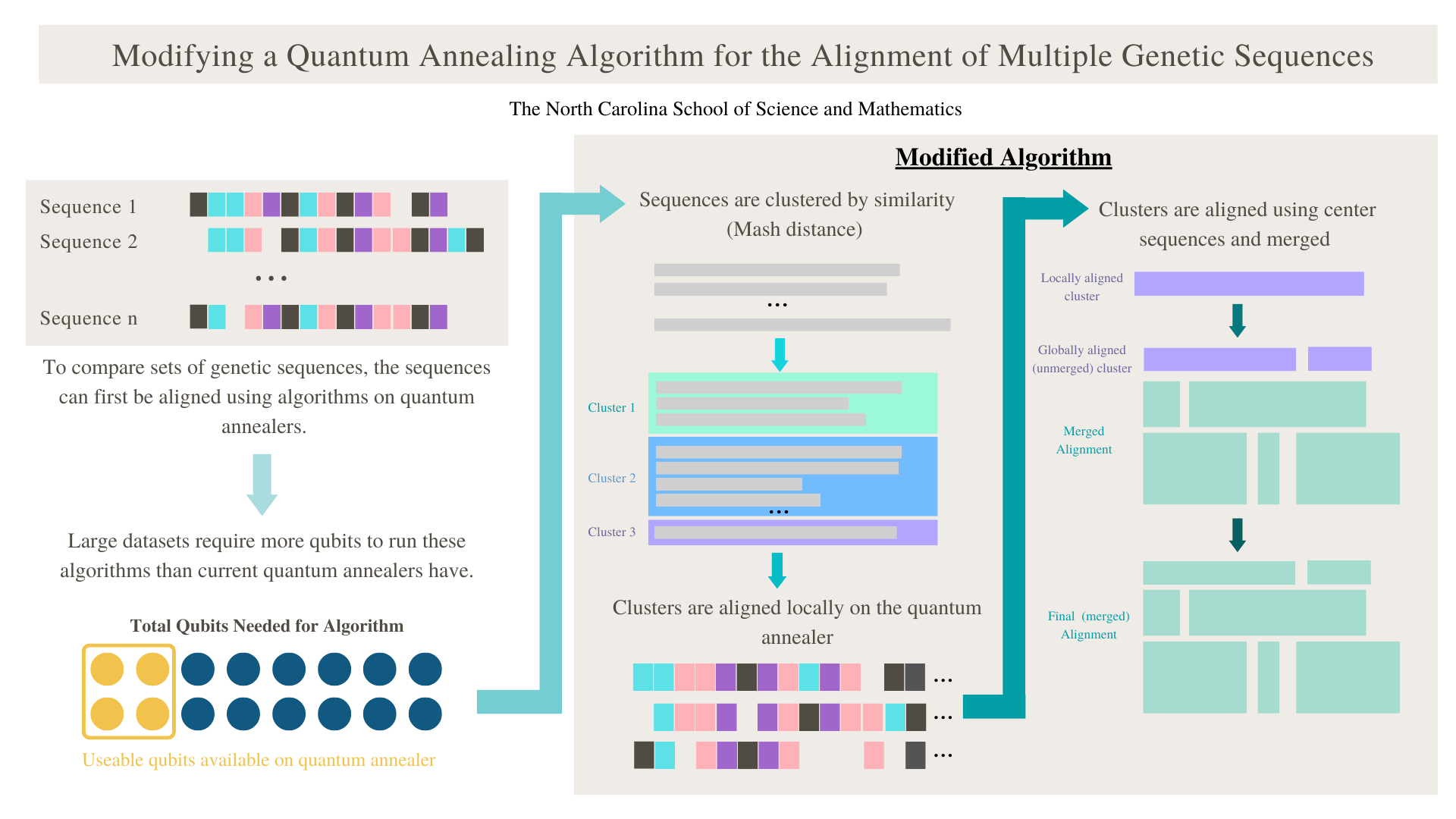}
    \centering
    \caption{Visual obserview of MAQ algorithm approach (created by author).}
    \label{Visual Abstract}
\end{figure}

We first determined the hardware on which the quantum algorithm could be run. This was used as a constraint to specify the algorithm body type. We then broke this project in three key parts: (1) Pre-processing, (2) Algorithm Body, and (3) Post-processing. These parts are defined by their function relative to the overarching algorithm, as outlined below and in Figure \ref{Visual Abstract}.

\begin{enumerate}
    \item The \textbf{Pre-processing [Key Modification]} part is the set of operations that reads in files and prepares the sequences for alignment.
    \begin{itemize}
        \item Read in sequences from FASTA file,
        \item Cluster sequences, and
        \item Convert sequence clusters into matrix.
    \end{itemize}
    \item The \textbf{Algorithm body} returns the alignments of given set of sequences. 
    \begin{itemize}
        \item Take in clusters and transform to form digestible by quantum solver and
        \item Align sequences per cluster.
    \end{itemize}
    \item The \textbf{Post-processing [Additional Modifications]} processes the results obtained from Parts 1 and 2. in order to produce a final output for the user.
    \begin{itemize}
        \item Interpret the annealing results,
        \item Merge locally aligned clusters with previous alignments, and
        \item Output final alignment.
    \end{itemize}
\end{enumerate}

\subsection{Hardware}
Thus, MAQ was run on the D-Wave Adiabatic Computing (Quantum Annealing) System, made accessible via the Leap integrated development environment (IDE). While other quantum annealers -- including those developed by the New Energy and Industrial Technology Development Organization \cite{nec_2018}, Ford Motor Cars, and Lockheed Martin \cite{obata_2022} -- exist, D-Wave annealers were selected due to their commercial availability and earlier establishment as a product available to the public \cite{dilmegani_2019}.

Simulations of this system were also accessible. The Leap IDE is a quantum cloud service run using Python. The D-Wave Solvers may also be used locally. Here, D-Wave's Ocean v.5.2.0 software development kit \cite{dwave-ocean-sdk} and dimod v.0.11.2 package \cite{dimod} was used, allowing the quantum annealing environment to be simulated on the local system's central processing unit (CPU).

\subsection{Pre-processing Development}
\subsubsection{Approaching Sequence Read-In And Storage}
Prior to aligning the sequences, we parsed the sequences in from an external file. We assumed that the data -- containing either protein or genetic sequences -- is contained on a single FASTA Formatted Sequence file. Using the Biopython v1.79 package, the sequences were stored as Sequence Record objects, containing key information on the sequence's identity \cite{cock2009biopython}. For the purposes of data storage and later processing, we assumed that every genetic sequence in the file had an unique identifier.

\subsubsection{Introducing the Novel Modification}
To implement the progressive alignment technique, we introduced a sequence clustering component to the pre-processing stage. To identify the clustering algorithm to accomplish this task, we first set the lowest possible bar. We observed the naive solution was not ideal. While the arbitrary assignment of sequences would not be costly, it would have come at the cost of the accuracy of the returned alignment. Thus a deliberate algorithm was sought for.

We took inspiration from the Feng-Doolittle progressive alignment approach \cite{feng_doolittle_1987}. The Feng-Doolittle algorithm uses classical computers to first group the clusters by similarity, then uses dynamic programming methods to merge the sequences following their local alignments \cite{feng_doolittle_1987}, producing an approximate MSA. We used a simplistic approach, which was in line with a similarly inspired hierarchical clustering algorithm developed in 1988 \cite{corpet_1988}.

More specifically, we adapted the ALFATClust algorithm and treated the clustering problem as a question of finding the nearest neighbor \cite{chiu_ong_2022}. It used the Leiden algorithm to localize each cluster, connecting "communities" of these clusters based on relative similarity \cite{traag_waltman_vaneck_2019}. This differs from the greedy approach taken by most existing software tools, which are reliant on a limited set of parameters (thereby producing not ideal alignments). 

To approximate the difference between sequences prior to clustering, ALFATClust uses the Mash (sample-based) technique \cite{chiu_ong_2022}. While preliminary studies have shown the alternate, unsupervised learning-based algorithms, such as MeShClust, are able to process these sequences more rapidly \cite{james_luczak_girgis_2018}, these algorithms return an unusually low number of clusters (with larger numbers of sequences per cluster) \cite{chiu_ong_2022}. This is contrary to one of the primary objectives of MAQ, which seeks to reduce the total spin usage once these datasets are passed into the quantum annealers. The ALFATClust method holds its own against other algorithms that do not employ the Mash heuristic, demonstrating its viability for selection for our purposes \cite{chiu_ong_2022}.

Following initial testing, it was revealed that ALFATClust occasionally returns clusters that contain a small number of sequences (e.g. a 2-sequence dataset), for which calls on a quantum annealer may be deemed unnecessary. To remedy this, we introduced a minimum cluster threshold size. If a cluster size did not meet the threshold, it would be appended to the next cluster, the entirety of which was then aligned locally.

To create a standard of comparison across each subsequent sequence, we introduced a function to identify the centers of each of the clusters. The center was defined as a singular sequence in the group with the lowest total distance when compared against all other sequences in the cluster. This center re-emerges in the post-processing stage to aid in the merging of cluster alignments.

The Mash v.1.14 package was used to conduct preliminary estimations on the distances between each of the sequences \cite{ondov_treangen_melsted_mallonee_bergman_koren_phillippy_2016}. The subsequent data was analyzed using the NumPy v.1.22.4 \cite{harris2020array}, SciPy v.1.8.1 \cite{2020SciPy-NMeth}, and Pandas v.1.4.2 packages \cite{reback2020pandas}. The clustering algorithm calls on the Leiden algorithm v.0.8.10 package \cite{traag_waltman_vaneck_2019} and Python igraph v.0.9.11 package \cite{igraph}.

\subsection{Main Algorithm}
Each cluster is then passed through the main algorithm, with the center from a previously aligned sequence appended to the cluster for later merging. To implement the MSA problem in the annealing algorithm, we defined the problem space, developing the Hamiltonian for the distance minimization problem with constraints. The algorithms were thusly based on this problem formulation. While selecting the algorithm for the body, we considered three sets of variables: appropriate use of the (1) objective, (2) weights and penalties, and (3) constraints. Quantum spin usage was a secondary driving factor. 

We defined (1) the objective to be the minimization the overall distance between the sequences. Thus, in constructing (2) the weights matrix, an effective method of comparison and storage must be used. Full penalties are applied in alignments where the elements in corresponding positions do not match. To avoid the insertion of unnecessary gaps, a partial penalty for these gaps are included. After the weights matrix in the sequences is found, (3) constraints may be applied. These constraints would be dependent on the approach.

We considered two potential approaches to problem formulation. We began by defining the parameters of the problem. When given a \(L\)-sized set of sequences with maximum sequence length \(N\), the naive solution is to use a systematic brute-force approach. Every element in each sequence will compared against every other element in all other sequences. In this case, every possible pairing of elements will require a corresponding spin value to be stored. This requires a system on the magnitude of \(O(N^L)\). This is by all means infeasible on current hardware, especially after gaps are inserted to account for element deletions or insertions (a biological phenomenon) \cite{lindvall_2019}. 

After further research, we determined a secondary, more effective approach. Oscar Lindvall proposed using the Column Alignment Formulation (CAF) approach to align the sequences (Figure 4). It may be visualized using a table with \(L\) rows and some \(C\) columns. Given some user-defined parameter \(G\), representing the maximum number of gaps that may be inserted into the sequence to shift corresponding sections, \(C\) can be set equal to \(N + G\). Every row represents a single sequence, and every column a single position. The goal, then, is to find the positions in every row where an element in the sequence can be placed, such that the number of differences per column is minimized. We assume \(G\) is significantly small relative to \(N\), the maximum length of the sequence. In this case, the number of spins that must be represented at any point (that is, the number of qubits needed) is within \(O(LN^2)\) \cite{lindvall_2019}, a reduction to manageable polynomial magnitude. 

\begin{figure}
    \includegraphics[width=\linewidth]{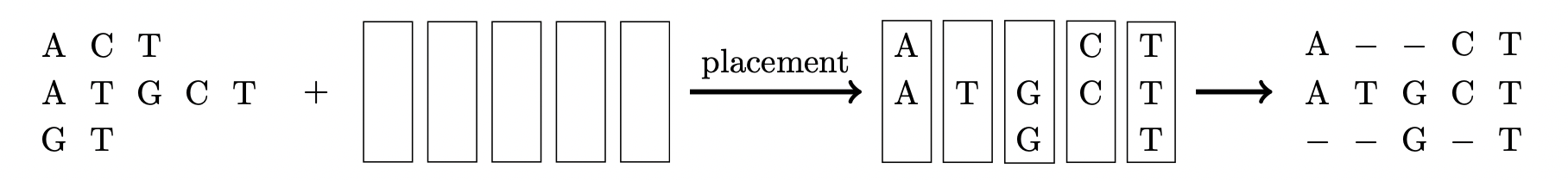}
    \caption{CAF aligns sequences by assigning the elements in each sequence a column and inserting gaps into any column spaces with no elements \cite{lindvall_2019}.}
    \label{CAF}
\end{figure}

We used the CAF approach proposed by Lindvall, applying some pre-defined penalty \(g\) for the insertion of empty spaces in the sequences \cite{lindvall_2019}. Per Lindvall's proposed algorithm, we constructed the matrix by comparing the sequences against one another each other, resulting in weights \(w_{(s1,n1,s2,n2)}\) for every pairing of elements \cite{lindvall_2019}. 

\subsection{Post-Processing Development: Dynamic Programming}
The output of the main algorithm is a matrix, where each row represents a sequence and each column a position \cite{lindvall_2019}. The first '1' in the row is where the first element in the corresponding sequence is placed, the second '1' is where the second element is placed, and so on. If a '0' exists in the matrix, then a gap has been inserted in that position. We authored a simple method to interpret these results and transform them into readable strings.

However, the process at this step is incomplete. Alignments have only been made for the individual clusters. Recall that the alignment contains the center of the previous cluster. Using comparisons between the gaps inserted in the center in this and the previous iteration of the algorithm, we dynamically merge the clusters together, such that after merging, the current cluster is immediately forgotten from the quantum annealer. 

\section{Results \& Complexity Analysis}
To properly analyze the preliminary results returned by this modified algorithm, we reiterate that the main goals of this project were to 

\begin{enumerate}
    \item Introduce classical-inspired heuristics to rudimentary quantum algorithms, and
    \item Reduce the spin usage per call of the quantum annealer.
\end{enumerate}

In order to approximate the effectiveness of the algorithm in achieving this end, we conduct a rough space complexity analysis of the key impacts of (1) the weights determination function, (2) the quantum-dependent component, and (3) the merging function. These three areas have been impacted most strongly by the modifications.

\subsection{Analysis of Weights Matrix Function}
Let us consider an input dataset of \(L\) sequences, with a maximum sequence length of \(N\) and \(G\) inserted gaps per sequence. We first consider the characteristics of the initial algorithm for comparison. The creation of the weights matrix is especially consuming, since it requires storage of the comparisons between every individual element in the dataset. Since every possible pair of elements in distinct sequences is compared, the space complexity may roughly be given by

\begin{equation} \label{eq1}
\begin{split}
    O(\frac{N!}{2!(N-2)!}\times \frac{L!}{2!(L-2)!}) \\
    \simeq O(\frac{N(N-1)}{2} \times \frac{L(L-1)}{2})\\
    \simeq O(N^2L^2)
\end{split}
\end{equation}

Let the clustering algorithm reduce the dataset to some number of clusters, such that the largest cluster has \(k\) sequences, where \(k << L\). The weights matrix determination function is then reapplied to this reduced sample size, resulting in a complexity of \[O(N^2k^2)\] per cluster. However, the weights matrix must be applied at most \(k\) times. Therefore, the overall complexity of the weights matrix is given by 

\begin{equation} \label{eq1.5}
    O(N^2k^2 \times \frac{L}{k}) \simeq O(N^2Lk).
\end{equation}

Equation \ref{eq1.5} presents a linear advantage over the initial weights matrix development requirements. However, this advantage is partially offset by the ALFATClust algorithm introduced during the pre-processing stage. Nevertheless, the ALFATClust's application of the Mash approximation for distance estimation cuts down significantly on the initial \(O(N^2L^2)\) space complexity \cite{chiu_ong_2022}.

\subsection{Analysis of Alignment Algorithm}
Secondly, we consider the spin usage during the sequence alignment on clusters. Spin usage is a quantitative approximation of the number of nodes that will be used on the quantum annealer during computation. Recall we seek to reduce this usage per call of the quantum annealer.

The use of the Column Alignment Formulation (CAF) method already introduces a significant reduction on possible spin usage. The spin values -- and resultant alignment -- is stored in some matrix, where the number of columns is equal to the sum of the length of the sequence and number of gaps 

\begin{equation} \label{eq1.9}
    C = N + G. 
\end{equation}

Thus, using Equation \ref{eq1.9}, we conclude the spin usage \(S\) is given by 

\begin{equation} \label{eq2}
    S=C\sum_{i=1}^{L}N_i
\end{equation}

where \(N_1... N_L\) are the lengths of the sequences \cite{lindvall_2019}. That is, the spin usage may be roughly described as

\begin{equation} \label{eq3}
\begin{split}
    S \in O(CLN) \\
    = O(LN(N + G) \\
    = O(L(N^2+NG) \\
    \simeq O(LN^2)
\end{split}
\end{equation}

where it is assumed \(G << L\) \cite{lindvall_2019}. 

We now consider the spin usage on a reduced number of sequences, given by \(k\). Following a similar line of reasoning, the spin usage when run on a single cluster may be given by 

\begin{equation} \label{eq4}
O(kN^2) << O(LN^2)
\end{equation}

Observe that the total spin usage (on the magnitude of \(O(kN^2 \times \frac{L}{k}) \simeq O(LN^2)\) is not representative of the maximum spin usage at a single point, as the quantum annealer is called \(L/k\) distinct times. It is worth noting that this rough complexity analysis treats the processing time of the inputs as a black box, thereby not accounting for the space or time needed to translate the input system onto the corresponding architecture (that is, node arrangement) for the annealer. This approach is nonetheless effective, as it indirectly implies the net node usage on the quantum annealer. Therefore, it follows from this reasoning that an approximate linear advantage is achieved in spin usage. 

\subsection{Analysis of Merging Function}
Lastly, we analyze the function that progressively merges the aligned clusters. The local alignments are stored in matrices containing the elements and gaps with their corresponding positions. These matrices have a maximum size \(O(kC)\), meaning the space complexity for \(n\) clusters may be described as 

\begin{equation} \label{eq5}
\begin{split}
    \sum_{i=0}^{n}k_iC\\
    \simeq O(LC)\\
    \simeq O(L(N+G))\\
    \simeq O(LN+LG)\\
    \simeq O(LN)
\end{split}
\end{equation}

These clusters are aligned locally. The center of the previously aligned cluster is included in the alignment of the new cluster. When merging, this center serves as the metric of comparison and the entire sequence is iterated through at least once, with a maximum length of \(N\), resulting in a minimum baseline runtime of \(O(N)\). Additionally, any gaps (\(G\)) that are inserted are then propagated throughout the remainder of the corresponding alignment (including through the compiled sequences in all previous alignments). Thus, over \(n\) clusters, the total running time is approximately

\begin{equation} \label{eq6}
\begin{split}
     O(N) + \sum_{j=0}^{n}kGj \\
     \simeq O(N) + O(kGn(n+1) \\
     \simeq O(N) + O(kGn^2)\\
     \simeq O(N + kGn^2)
\end{split}
\end{equation}

In the worst case scenario, to draw an upper bound on the runtime, \(n=L\) and \(k = 1\). Then, the worst case runtime is roughly

\begin{equation} \label{eq7}
    O(N + GL^2)
\end{equation}

\section{Testing the Algorithm}
The developed algorithm, named MAQ, was run on a small, sample dataset for comparison (Table \ref{maq_results}). Throughout the development process, the algorithm was repeatedly tested on this reduced dataset. Each sequence in the dataset was a derivation of some "base" sequence that represented some accepted sequence, along with an identical sequence "control" that ensured the most basic alignment (of the same sequences) could be achieved. Each subsequent sequence then contained at least one fundamental mutation that may occur in generic sequences (e.g. insertion, deletion, or point mutations). The sequences are identified in Table \ref{maq_results} accordingly. When run on ALFATClust, the dataset is clustered into three distinct sets of sequences, making it ideal to test the clustering-based MAQ algorithm.

\begin{table}[htb]
\centering
\caption{Alignment returned by MAQ algorithm using sample dataset (created by author).}
\begin{tabular}{llllllllll}
\textbf{ID} & \multicolumn{9}{l}{\textbf{Sequence Alignment}} \\
Base & - & N & V & R & L & M & L & R & L \\
Control & - & N & V & R & L & M & L & R & L \\
Insertion & M & N & V & R & L & M & L & R & L \\
Deletion & - & N & - & R & L & M & L & R & L \\
Point & - & N & V & M & L & R & L & N & L \\
InsertionAndDeletion & M & N & V & R & L & - & R & - & L
\end{tabular}
\label{maq_results}
\end{table}

\begin{table}[htb]
\centering
\caption{Alignment returned by Oscar Lindvall's algorithm \cite{lindvall_2019} using sample dataset (created by author).}
\begin{tabular}{llllllllll}
\textbf{ID} & \multicolumn{9}{l}{\textbf{Sequence Alignment}} \\
Base & N & V & - & R & L & M & L & R & L \\
Control & N & - & V & R & L & M & L & R & L \\
Insertion & M & N & V & R & L & M & L & R & L \\
Deletion & N & - & R & L & - & M & L & R & L \\
Point & N & - & V & M & L & R & L & N & L \\
InsertionAndDeletion & M & N & V & - & R & - & L & R & L
\end{tabular}
\end{table}

\begin{table}[htb]
\centering
\caption{Alignment returned by Kalign \cite{lassmann_sonnhammer_2005} using sample dataset (created by author).}
\begin{tabular}{llllllllll}
\textbf{ID} & \multicolumn{9}{l}{\textbf{Sequence Alignment}} \\
Base & - & N & V & R & L & M & L & R & L \\
Control & - & N & V & R & L & M & L & R & L \\
Insertion & M & N & V & R & L & M & L & R & L \\
Deletion & - & - & N & R & L & M & L & R & L \\
Point & - & N & V & M & L & R & L & N & L \\
InsertionAndDeletion & M & N & V & R & L & R & L & - & -
\end{tabular}
\end{table}

\subsection{Metrics of Comparison}
We firstly define the metrics used to compare these three MSA tools. We quantify the effectiveness of the algorithm by considering the alignment's deviation from the norm. The analysis is considered by column (following the CAF methodology), with pairwise comparisons conducted. In other words, we use a sum-of-pairs scoring method. For every pair of elements that differ in the same column, the total score for the alignment is incremented by \(+1\), although differences between base pairs or amino acids and gaps will have no penalty (an adjustable parameter during the development of the problem space). An ideal alignment will have a total score of 0. The greater the alignment score, the less effective the alignment. 

We now define this alignment score formally. Let us label the sequences in the final alignment from \(\{s_0, s_1, ...., s_L\}\), organized in a matrix containing \(C\) columns and \(L\) sequences. Note these aligned sequences include any gaps inserted after the dataset is processed using the alignment algorithm. Then, construct a new matrix \(A\) with dimensions \(C \times L \times L\), where the element \(a_{c, i, j}\in A\) equals 1 if the \(c\)th element of sequences \(s_i\) and \(s_j\) are not equivalent and are not gaps and 1 otherwise. Then, the alignment score is defined as 

\begin{equation} \label{eq8}
    \sum_{c=1}^{C}\sum_{i=1}^{L}\sum_{j=i}^{L}A_{c,i,j}.
\end{equation}

For example, consider the set of sequences \({AT, T}\). An example alignment may be seen in Table \ref{sample_scoring_alignment}. Observe that the first column has 3 pairs of alignments that do not match. The pair \((A, T)\) has weight \(+1\), while the pairs \({(A, -), (-, T)}\) do not match but contain gaps, so these differences are weighted at \(0\). Observe that the second column does not contain any pairwise differences. Using Equation \ref{eq8}, we find the score of the alignment in Table \ref{sample_scoring_alignment} is \(1\). 

\begin{table}[htb]
\centering
\caption{Sample genetic sequence alignment, with a resultant alignment score of 1 (created by author).}
\begin{tabular}{ll}
A & T \\
- & T \\
T & T
\end{tabular}
\label{sample_scoring_alignment}
\end{table}

\subsection{Comparing with Existing Algorithms}
We conducted preliminary tests on MAQ and compared the results obtained against results from two other algorithms: the unmodified Lindvall algorithm and a classical algorithm that uses similar progressive techniques. Much like how MAQ clusters sequences into local groups prior to alignment, Kalign focuses on alignments in local regions \cite{madeira_pearce_tivey_basutkar_lee_edbali_madhusoodanan_kolesnikov_lopez_2022}, employing a heuristic version of the Wu-Manber string (sequence) alignment algorithm. Kalign was shown to be significantly more accurate than other methods on large datasets, especially when compared against popular methods, such as Balibase and Prefab \cite{lassmann_sonnhammer_2005}. The algorithm was an estimated 10 times faster than ClustalW, an algorithm that makes use of tree-like data structures (arguably a more sophisticated form of clustering) to align the sequences \cite{thompson_higgins_gibson_1994}. 

After the test dataset of sequences (as seen in Table \ref{maq_results}) was aligned on the three algorithms (MAQ, Lindvall's, and Kalign), the alignment scores were calculated using Equation \ref{eq8} and organized in Table \ref{multiple_algorithm_comparison}. 

\begin{table}[htb]
\centering
\caption{MAQ is able to return an alignment with competitive alignment scores on relatively small sets of sequences (created by author).}
\label{multiple_algorithm_comparison}
\begin{tabular}{l|lllllllll|l}
\multirow{2}{*}{\textbf{Algorithm}} & \multicolumn{9}{l|}{\textbf{Alignment Score by Column}} & \multirow{2}{*}{\textbf{Total}} \\ \cline{2-10}
 & \textbf{1} & \textbf{2} & \textbf{3} & \textbf{4} & \textbf{5} & \textbf{6} & \textbf{7} & \textbf{8} & \textbf{9} &  \\ \hline
MAQ & 0 & 0 & 0 & 5 & 0 & 4 & 5 & 4 & 0 & 18 \\
Lindvall & 8 & 2 & 4 & 7 & 4 & 4 & 0 & 5 & 0 & 34 \\
Kalign & 0 & 0 & 5 & 5 & 0 & 8 & 0 & 4 & 0 & 22
\end{tabular}
\end{table}

\section{Major Conclusions}
The world of bioinformatics shapes societal responses to disease. A significant part of this understanding arises from pattern identification, which may be used to find information to predict how patients may respond to various diseases or treatments. This poses a series of sequence-based problems that are solvable on algorithms. Among these, MSA plays a significant role. After all, comparison of large sets of genetic or protein sequences is reliant on the assurance that these sets have been aligned in a logical manner. In spite of its relevance, the problem is NP-complete, which speaks to the need for the development of algorithms that are capable of stepping beyond the 0’s and 1’s of today’s classical computers. Our developed algorithm, MAQ, is one step in such this direction. 

The application of quantum computing to problems is not new \cite{marx_2021}. Over the years, algorithms for tasks such as genetic sequencing and protein structure prediction have been proposed \cite{marx_2021}. However, many are heavily restricted by spin usage and the relatively new nature of the field. MAQ introduces a classical-inspired approach reduces the spin usage per call of the quantum annealer.

The algorithm first clusters the sequences using ALFATClust \cite{chiu_ong_2022}. The reduced sequence sets are then compared and aligned on the main algorithm, modified from Oscar Lindvall's approach \cite{lindvall_2019}. The resultant alignments are then dynamically merged based on the relative spacing of the center sequences of each cluster. The final, progressively aligned alignment is then returned to the user.

A linear advantage of \(O(L/k)\), given \(L\) total sequences and \(k\) clusters, is achieved in the reduction of spin usage per call of the quantum computer (Equation \ref{eq1.5}). However, added complexity due to the addition of the clustering step and repetitive calls to the quantum annealer adds to the overarching running time. Nonetheless, the spin usage of each single call on quantum annealers has been reduced. This allows for the adjustment of large datasets for current quantum hardware that has yet to be able to handle significant space usage without significant loss of information.

Furthermore, when run on a test dataset, MAQ was shown to be comparable to existing MSA algorithms, including Lindvall's initial algorithm and Kalign. For this specific dataset, MAQ performed better, with an alignment score of 18, relative to the scores of 34 and 22 for Lindvall's algorithm and Kalign, respectively. Thus, it is comparable to existing algorithms.

\section{Discussion and Wider Applications}
One must caution that the advantage achieved by MAQ is dependent on the characteristics of the data. The viability of a clustering-based method may be determined using the rank of the set of sequences (that is, how similar the sequences are to each other, where lower rank suggests larger similarity). The lower the rank of the set, the more likely the results will resemble that of Lindvall's algorithm, since the number of clusters is reduced. MAQ assumes there exists sufficient distinctions between each sequence in the set such that they may be clustered into a reasonable number of subsets. In other words, there is moderate variability between the sequences. In the case all of the sequences are nearly identical (say, with an estimated similarity of \(>0.99\), the clustering may be deemed ineffective. Consider the alternative extreme. In the case the sequences have unusually high rank (where the variability between the sequences is high), the number of clusters will be close to the initial number of sequences, and the impact of the clustering algorithm will be called into question. One may argue the sequences in these extreme cases should instead be grouped by the order it is read in from the file. This would consume fewer resources. 

Future research is needed to quantify the actual effectiveness of clustering prior to alignment. This is especially important since the clustering algorithm is costly, as it is itself tackling a NP-hard problem \cite{mahajan_nimbhorkar_varadarajan_2012}. further study may reveal a definitive response on whether the cost of clustering the sequences exceeds the benefits of an improvement in alignment when compared against, for instance, alignment of random groupings of sequences.

Furthermore, future MAQ versions may explore other algorithms, including those that use K-means clustering, where the number of clusters is predefined. Then, the approximate reduction of the spin usage per call on the quantum annealers may be approximated with greater certainty. Granted, although the number of cluster will be guaranteed (including for sets with low rank), the size of these clusters will still be dependent on user parameters.

This algorithm deserves further revisitation. Tackling MAQ as three distinct components that funnel into one another presents an opportunity for improvement. Additional research is needed to investigate approaches to consolidating sequence clustering and alignment, especially with regards to the creation of the weights matrix (a costly process). For example, the pairwise distance of sequences is first estimated using the Mash heuristic during the clustering pre-processing phase. The pairwise comparisons are then completed a second time while creating the problem space the quantum annealer will solve (although the exact mechanics differ). Thus, a standalone clustering algorithm may not be the best integration into MAQ. Rather, future versions of MAQ may look to consolidate these pairwise comparisons to reduce overall iterations through the sequences. Alternative approaches should also be studied.

Additionally, when run on small datasets, the quantum annealing-based algorithms may regularly return different results. This is likely the result of multiple "lowest energy state" configurations. In MAQ, these differences may be propagated across the clusters, magnifying minor decisions early on in the alignment process between mathematically-identical alignment states. For each cluster aligned, there is no guarantee that the arbitrarily chosen state will result in the lowest alignment score across the total alignment. It merely guarantees a low alignment score locally. In order words, the dynamic merging process assumes all previous alignments are ideal, an assumption that does not always hold true. Despite this, this characteristic of the algorithm may be harnessed as an advantage. For example, rather than returning a single plausible alignment, several alignments -- one corresponding to each combination of the ideal, local solutions -- may instead be simultaneously compared by the algorithm. This may open the door for a more accurate final solution to be returned. Further research is needed to explore alignment algorithms that may make the most of the existence of a set of plausible local alignment results.

MAQ demonstrates the viability of quantum computing as a supporting system for studies into computational biology. This is a part of the wider driving force that dictates the possible paths of research development. After all, MSA is just one of many optimization problems in bioinformatics. Genetic engineering and sequencing, for example, are heavily reliant on the capabilities of existing technology. These capabilities are defined by the accuracy and accessibility of these tools. As the accessibility of quantum computers increases, a rising number of algorithms -- including MAQ -- are bridging the gap between quantum computing and other areas.

These identified areas have the potential to impact millions of human lives. Chief among them are epidemiological and phenological studies. In particular, comparison of these sequences permits for a stronger understanding of the human genome. More rapid sequencing tools will help translate compiled genomic data into medically useful information \cite{gonzaga-jauregui_lupski_gibbs_2012}. This includes a better approach to treatment response prediction -- including chemotherapy -- and phenology determination of disease strains. Through extensive multiple sequence analysis (made possible through alignment), medical professionals' understanding of the genetic patterns corresponding to phenotypical characteristics may be expanded. These developments have the potential to impact the 33.4 million individuals who pass through the US hospital system annually \cite{ahahospitalstatistics_2022}, along with the countless others who use any form of healthcare service. In order to achieve this, however, refinement of the quantum annealing process and algorithms must be conducted. As problem sizes continue to grow and the need for algorithms with lower space complexity and runtimes continues, heuristics such as that taken by MAQ will continue to emerge, marking this as an area of strong potential, worthy of further research.

\section{Student Reflection}
MAQ was the result of a 9-month student research project I (the author) conducted. The investigative project explored the plausibility of applying quantum computing as a tool. In particular, I focused on addressing current limitations of the quantum hardware. However, arriving at this focus involved a rather indirect path consisting of a series of decisions. 

My initial research had led me into a more abstract form of string alignment. This pure mathematics problem approached the situation via graph theory and employed techniques beyond the scope of this paper. I had initially begun with the intention of applying my previous understanding of quantum-inspired and quantum computing algorithms to the project. Yet these purely theoretical subjects felt disconnected from ongoing problems in the world, and I struggled to identify a path forwards. 

Over time, as the research plan began to solidify, I encountered an increasing number of applications for these algorithms. The puzzle pieces began to fall into place as I read about the application of MSA algorithms to genetic sequencing. I found I was revisiting a subject that had fascinated me years prior, and my appreciation for interdisciplinary studies grew.

This played a role in reshaping my long-term plans for study. In particular, my focus transitioned from pure mathematics and theoretical computer science to computational biology. While the two former fields are still on my radar as fields of interest, I recognize computational biology will likely play a larger role in the direction I take for future endeavors. Following conversations with a number of current graduate students, professors, and researchers in the field, I hope to go into and remain in research and academia following my college and (ideally) graduate studies.

Nevertheless, I recognize this research project is merely a small glimpse of what is plausible in the realms of bioinformatics and quantum computing. Even as I have gained a stronger understanding of algorithmic thinking, implementing quantum annealing, and the mathematics surrounding the fields, I realize I have much more left to learn. I have no intention of stopping my curiosity, and I hope to continue to expand my understanding of what is possible over the next few decades.

\section{Acknowledgements}
Thank you to Mr. Robert Gotwals for providing feedback and guidance throughout the research process.


\bibliographystyle{unsrt}  
\bibliography{references}

\end{document}